\documentclass[12pt]{article}
\usepackage{graphicx}
 \usepackage{longtable}
\usepackage{multirow}
\relax
\usepackage{amssymb}

\newcommand{\bo}{{\bar o}}

\def\bo{{\raise.15ex\hbox{\large$\Box$}}}               

\def\face{{\raise.2ex\hbox{$\displaystyle \bigodot$}\mskip-2.2mu \llap {$\ddot
        \smile$}}}                                      

\def\Zbf{{\bf Z}}

\def\leftrightarrowfill{$\mathsurround=0pt \mathord\leftarrow \mkern-6mu
        \cleaders\hbox{$\mkern-2mu \mathord- \mkern-2mu$}\hfill
        \mkern-6mu \mathord\rightarrow$}       
\def\dvec#1{\vbox{\ialign{##\crcr
        \leftrightarrowfill\crcr\noalign{\kern-1pt\nointerlineskip}
        $\hfil\displaystyle{#1}\hfil$\crcr}}}           

\def\beq{\begin{equation}}
\def\eeq{\end{equation}}

\def\beqx{\begin{displaymath}}
\def\eeqx{\end{displaymath}}

\def\beql{\begin{eqnarray}}
\def\eeql{\end{eqnarray}}

\newcommand{\bea}{\begin{eqnarray}}
\newcommand{\eea}{\end{eqnarray}}

\def\[{\left [}
\def\]{\right ]}
\def\({\left (}
\def\){\right )}

\def\+{\oplus}

\begin{document}

\hbox{\hskip 12cm NIKHEF/2010-45  \hfil}
\hbox{\hskip 12cm IFF-FM-2010/03  \hfil}
\hbox{\hskip 12cm December 2010  \hfil}

\vskip .5in

\begin{center}
{\Huge \bf Asymmetric Gepner Models} \\  \vskip .2truecm {\Large  \bf III. B-L Lifting\footnote{This paper is dedicated to the memory of Max Kreuzer.}}

\vspace*{.4in}
{ B. Gato-Rivera}$^{a,b,}\footnote{Also known as B. Gato}$
{and A.N. Schellekens}$^{a,b,c}$
\\
\vskip .2in

${ }^a$ {\em NIKHEF Theory Group, Kruislaan 409, \\
1098 SJ Amsterdam, The Netherlands} \\

\vskip .2in

${ }^b$ {\em Instituto de F\'\i sica Fundamental, CSIC, \\
Serrano 123, Madrid 28006, Spain} \\

\vskip .2in

${ }^c$ {\em IMAPP, Radboud Universiteit,  Nijmegen, The Netherlands}

\end{center}

\vspace*{0.3in}
{\small
In the same spirit as heterotic weight lifting, B-L lifting is a way of replacing the superfluous and ubiquitous $U(1)_{B-L}$ with
something else with the same modular properties, but different conformal weights and ground state dimensions. This method
works in principle for all variants of $(2,2)$ constructions, such as orbifolds, Calabi-Yau manifolds, free bosons and fermions and
Gepner models, since it only modifies the universal $SO(10) \times E_8$ part of the CFT. However, it can only yield  chiral spectra if
the ``internal" sector of the theory provides a simple current of order 5. Here we apply this new method to Gepner models. Including
exceptional invariants, 86  of them have the required order 5 simple current, and 69 of these yield chiral spectra. Three family spectra
occur abundantly. 
}


\vskip 1in

\noindent
\newpage

\section{Introduction}

In a previous paper \cite{GatoRivera:2009yt} we introduced a new way of constructing heterotic strings we called
``heterotic weight lifting". This construction is part of a presumably huge class of constructions based on the general idea of
replacing building blocks  of the left-moving bosonic sector of the heterotic string by something else in the same representation
of the modular group. This manifestly preserves modular invariance, and since the fermionic  sector is not changed, world-sheet
supersymmetry is preserved as well. Space-time supersymmetry is, as always, optional. This method is especially useful if one starts with
$(2,2)$ CFTs. In that case, the canonical solution to modular invariance is to make the left and right sector identical, and map the
left one to a bosonic string using the bosonic string map \cite{LLS}. Weight lifting allows us to move further away from the $(2,2)$ CFTs
and into the genuine $(0,2)$ landscape, by removing the superfluous $N=2$ world-sheet supersymmetry in the bosonic sector.
The natural area of application of this method is in constructions based on {\it interacting} CFTs, in particular ``Gepner models"  \cite{Gepner:1987qi}, where $(2,2)$ spectra are the
canonical starting point. For {\it free} CFTs other methods to construct $(0,2)$ spectra are available, and have
already been studied extensively, see {\it e.g.}  \cite{Assel:2010wj,Lebedev:2006kn}
and references therein.

The benefits of moving towards genuine (0,2) models have been demonstrated in  \cite{GatoRivera:2010gv} and \cite{GatoRivera:2010xn}. In these papers
standard Gepner models  and lifted Gepner models were studied respectively. It turns out that the case of 
interest for phenomenology, three families, is rare in standard Gepner models but very common among
lifted Gepner models. It was already observed in the late eighties of last century 
(see {\it e.g.} the extensive studies published in  \cite{Schellekens:1989wx,Schellekens:web,Fuchs:1989yv}) that the number of families
in simple current modified (2,2) models is usually a multiple of six, occasionally two or four, and very rarely three. This was confirmed in  \cite{GatoRivera:2010gv}
even when the canonical $SO(10)$ gauge group is broken to $SU(3)\times SU(2) \times U(1)^2$. But if we do the same analysis
with lifted Gepner models, one finds an abundance of distinct three family models, which are not related to each other in any known way.
Indeed, since they originate from different $N=2$ tensor products, one would not expect them to lie in the same moduli space, just
as distinct standard  Gepner tensor products are expected to belong to different Calabi-Yau moduli spaces. A different class of examples 
of $(0,2)$ interacting CFT heterotic spectra with three families was presented in
\cite{Blumenhagen:1996vu}.

\section{B-L Lifting}

Heterotic weight lifting replaces one (or more) minimal model factors and the $E_8$ affine Lie algebra by an isomorphic CFT. 
There is another obvious candidate for such a replacement: the $U(1)$ that is the remainder of $SU(3)\times SU(2) \times U(1)$ within
$SO(10)$. If the standard model embedding is the canonical group-theoretical one, this $U(1)$ (more precisely, a suitable linear combination of the extra $U(1)$ and $Y)$ corresponds to $B-L$. However, in many
spectra we have studied in \cite{GatoRivera:2010gv} and \cite{GatoRivera:2010xn} it couples differently to quarks and leptons
than $B-L$, and it may even be anomalous\rlap.\footnote{Note that when the affine Lie algebra $SO(10)$
is broken to $SU(3)\times SU(2)\times U(1)$ in string theory, the representations do not necessarily
decompose according to the group-theoretical decomposition of the $(16)$ of $SO(10)$.}
Nevertheless, for simplicity we will refer to this extra $U(1)$ factor as ``B-L". The idea
of  ``B-L lifting" is to replace this $U(1)$ (which has twenty characters, and will be denoted $U(1)_{20}$)
by another CFT whose characters have the same modular transformations. We must also
match the central charge. This can be done by either finding another CFT at $c=1$ with the same number of characters and
the same modular transformations matrices $S$ and $T$, or by using up the $E_8$ factor, so that we consider $c=9$. Since the
$c=1$ models are all known, it follows that the former is not an option. On the other hand, finding a $c=9$ CFT with twenty characters
is easy. The fact that $S$ must be the same tells us that it must have the same fusion rules as $U(1)_{20}$, and in particular a 
$\Zbf_{20}$ simple current symmetry. An obvious candidate is the product of affine level one algebras $A_4 \times D_5$, and by simple inspection we find that indeed
this is a solution. 

It is, however, not the only solution. To find all solutions, one can proceed as follows. First
we determine a {\it complement} of $U(1)_{20}$. This is a CFT with 20 characters and
a set of modular transformation matrices $S$ and $T$ that are the complex conjugates of those
of $U(1)_{20}$. A simple realization of such a complement 
is the coset ${\bf W}=E_8/U(1)_{20}$.
It can be written as ${\bf W}=SU(5)\times SO(6)$. This follows immediately from the group theory
embedding chain underlying the standard model
embedding in $E_8$: $E_8 \subset SO(6) \times SO(10) \subset SO(6) \times SU(5) \times U(1)_{20}$.
The CFT ${\bf X}$ we are looking
for has $c=9$ and is isomorphic to $U(1)_{20}$. Then its characters can be combined with those of 
${\bf W}$ to a meromorphic\footnote{This is a CFT with just one representation and hence just one
Virasoro character, that therefore is modular invariant by itself. We make use of meromorphic CFTs here
because for $c=8$ and $c=16$ they have been classified completely.}
$c=16$ CFT. There are only two of them, $E_8\times E_8$ and $SO(32)/\Zbf_2$, the ones used in the ten-dimensional heterotic string. 
Hence to find all solutions for ${\bf X}$ we can look for all embeddings of the affine algebra $SU(5)\times SO(6)$ at level 1 in 
$E_8\times E_8$ and $SO(32)/\Zbf_2$. There are just two embeddings in $E_8 \times E_8$: One can either embed both factors,
$SU(5)$ and $SO(6)$, in the same $E_8$ or in a different $E_8$. The former leads to a commutant $U(1)_{20}\times E_8$, the
standard $B-L$ solution, and the latter leads to a commutant $SU(5)\times SO(10)$, the lifted solution described above. 

In $SO(32)$ we can find another solution
\begin{eqnarray*}
SO(32)&\supset& SO(6)\times SO(26) \\
&\supset& SO(6)\times SO(10) \times SO(16) \\
&\supset& SO(6)\times SU(5) \times U(1)_{20} \times SO(16) 
\end{eqnarray*}
The factors that are left after removing $SO(6)\times SU(5)$ define the lift CFT ${\bf X}$, but 
these group theory embedding chains do not provide all information required to define them. The reason is that there are in general
some additional chiral algebra extensions, originating from the embedding and from the $SO(32)$ spinor current extension
that makes it into a meromorphic CFT. If some of these currents have spin 1 the groups appearing in ${\bf X}$ here are enlarged. 
This may lead to the suspicion that this  solution is in fact identical to the canonical $U(1)_{20}\times E_8$ solution. 
However, that cannot be true, since $SO(32)$ does not contain $E_8$. 
Indeed, in this case the algebra $U(1)_{20} \times SO(16)$ is extended by the current $(10,v)$, where ``10" denote the
charge 10 self-conjugate field in $U_{20}$ and $v$ is the  $SO(16)$ vector. 
So this solution is {\it not} identical to $U(1)_{20} \times E_8$.

To summarize: there are three distinct lifts ${\bf X}$ of the ``B-L" factor $U(1)_{20}$, namely
\begin{itemize}
\item{${\bf X}=U(1)_{20}\times E_8$, the trivial lift.}
\item{${\bf X}=SU(5)\times SO(10)$.}
\item{${\bf X}=U(1)_{20}\times SO(16)$ extended by $(10,v)$.}
\end{itemize}
There is nothing unusual about having more than one non-trivial lift;
note that also heterotic weight lifting for $N=2$ models is not unique for certain values of the ``level" $k$.

 In Table \ref{TableLift} we indicate to which $SU(5)\times SO(10)$ or $SO(16)\times U(1)$ representation each $B-L$ charge is
lifted, and by how much the conformal weight is changed. In both cases charge 7 is massive both before and after the lift. 
In comparison to heterotic weight lifting of $N=2$ minimal models, the term ``lift" is a bit inappropriate here, because as many states move up as
 move down.

\begin{table}[h]
\begin{center}
\vskip .7truecm
\begin{tabular}{|c|c|c|c|c|}
\hline
\hline
$B-L$ charge & $SU(5)\times SO(10)$ repr. & Lift  & $SO(16) \times U(1)$ repr. & lift \\
\hline
0 & (1,1)& 0  &  (1,0)      &   0     \\
1 & (5,16)& $+1$  &   $(\overline{128},1)$     & $+1$       \\
2 & (10,10)& $+1$  &   (1,2)     &   0    \\
3 & ($\overline{10}$,$\overline{16}$)& $+1$  &  $(\overline{128},3)$       &    $+1$     \\
4 & ($\bar 5$,1)& 0  &  (1,4)      &     0   \\
5 & (1,16)& 0  &  $(128,5)+(\overline{128},-5)$      &  0      \\
6 & (5,10)& 0  &    $(1,6)+(16,-4)$    &   0     \\
7 & (10,$\overline{16}$)& 0  &  (128,-3)      &   0     \\
8 & ($\overline{10}$,1)& $-1$  &  (16,-2)      & $-1$       \\
9 & ($\bar 5$,16)& $-1$  &  $(\overline{128},1)  $    &      $-1$  \\
10 & (1,10)& $-2$  &  (16,0)      &$-2$\\
 \hline
\end{tabular}
\vskip .7truecm
\caption{\small Lifts of all $B-L$ charges, up to charge conjugation.}
\label{TableLift}
\end{center}
\end{table}

We will first consider the first non-trivial lift on the list.  
The effect of this replacement is that a spectrum that previously had a $SU(3)\times SU(2) \times U(1)_Y \times U(1)_{20} \times E_8$ 
gauge symmetry is mapped to one with a $SU(3)\times SU(2) \times U(1)_Y \times SU(5) \times SO(10)$ gauge symmetry (as explained
in \cite{GatoRivera:2010gv}, $U(1)_Y$ is a $U(1)$ with 30 primaries, and will often be denoted as $U(1)_{30}$.)
We are
ignoring here the extra gauge symmetry coming from the internal sector. In the case of Gepner models, this is usually $U(1)^n$ where
$n$ is the number of factors. Occasionally this symmetry may be extended, or it may be part of an extension of the standard model (as
in the well-known case of an extension to $E_6$). Note that this lifting takes place entirely within the universal $SO(10)\times E_8$ 
factor that is present in any $(2,2)$ model. Hence one may modify any $(2,2)$ partition function in this manner. Therefore {\it a priori}
this method can be applied to $(2,2)$ models obtained using free bosons and/or free fermions, Calabi-Yau constructions, orbifolds
and Gepner models. 

However, one does not get something interesting in all cases. 
The reason is that in the fermionic sector we have to extend  $SU(3)\times SU(2) \times U(1)_Y \times U(1)_{20}$ back to $SO(10)$ in order
to use the inverse of the bosonic string map to get NSR fermions. To preserve modular invariance we have to perform a modularly isomorphic extension in the bosonic sector. Since the character bases are isomorphic there is a canonical choice. We can simply extend
the left chiral algebra by the lift of the simple current we use in the right chiral algebra. That current is $(3,2,1,4)$, where the last two
entries are the $U(1)$ charges (we denote representations of non-abelian groups by their dimension, and of abelian groups by
their charge, in units of the smallest charge). 
In the lifted theory this is mapped to the $SU(3) \times SU(2) \times U(1)_{30} \times SU(5) \times SO(10)$
representation $(3,2,1,5,1)$. This is a simple current of order 30. 
Its fifth power (which of course has order six) extends $SU(3) \times SU(2) \times U(1)_{30}$ to
$SU(5)$, the standard GUT extension. In this extended $SU(5)\times SU(5)\times SO(10)$ theory the current is $(5, 5,1)$. If we extend
the algebra by it, we get $E_8 \times SO(10)$. 

It is now clear what is happening. In both cases, we are embedding the standard model in the gauge group 
$SO(10)\times E_8$ that is always present in straightforward $(2,2)$ compactifications. However, the difference is that
instead of embedding the standard model in the canonical $SO(10)$, we have embedded
it into an $SO(10)$ subgroup of the extra $E_8$ factor.  Since $E_8$ only has non-chiral representations, there is no chance of getting a chiral
spectrum out of such a construction, or so it seems. 

But there is a way around  this obstacle. If one allows more general MIPFs (modular invariant partition functions) there is no reason why
the modular image of  the fermionic sector current $(3,2,1,4)$, that is the current  $(3,2,1,5,1)$,  should lie entirely within the $SO(10)\times E_8$ part of the
bosonic sector. If it does not, this current or some of its powers can be mapped to a higher spin current ({\it i.e.} with conformal weight larger than 1), which does not
lead to an extension of the gauge symmetry. Then any algebra in the subgroup chain between the unextended theory and the full $E_8$
extension can be realized. This leads to the same list of eight algebras presented in \cite{GatoRivera:2010gv}, but with $U(1)_{20}$
replaced by $SU(5)$ (since the current $(3,2,1,5,1)$ is trivial in the last factor, $SO(10)$, this group is not involved in any standard model
extension and remains always as a separate factor).
We then get the possibilities shown in Table \ref{TableG}.

\renewcommand{\arraystretch}{1.2}
{\small
\begin{table}[h]
\begin{center}
\vskip .7truecm
\begin{tabular}{|c|c|c||c|c|c|c|}
\hline
\hline
Nr. & Unlifted & Current & Order & Gauge group    & Charge quantum \\
\hline
0 &\small SM, Q=1/6& \small $(1,1,0,1,1)$ & $1$ & {\footnotesize  $SU(3)\times SU(2) \times U(1) \times SU(5)$} &    $ \frac16$  \\
1 &\small SM, Q=1/3 &\small $(1,2,15,1,1)$ & $2$ &  {\footnotesize $SU(3)\times SU(2)\times U(1)\times SU(5)$}  &    $ \frac13$  \\
2 &\small SM, Q=1/2 &\small $(3,1,10,1,1)$& $3$ &  {\footnotesize $SU(3)\times SU(2)\times U(1)\times SU(5)$}  &    $ \frac12$  \\
3 &\small LR, Q=1/6 &\small $(1,1,6, 5,1)$ & $5$ & {\footnotesize $SU(3)\times SU(2) \times SU(6)$}     & $ \frac16$  \\
4 &\small SU(5) GUT&\small $(\bar 3,2,5,1,1)$ & $6$ & {\footnotesize  $SU(5) \times SU(5)$}     & 1  \\
5 &\small LR, Q=1/3 &\small $(1,2,3,\overline{10},1)$ & $10$ &  {\footnotesize $SU(3)\times E_6$}  &    $ \frac13$  \\
6 &\small Pati-Salam &\small $(\bar 3,0,2,{10},1)$ & $15$ & {\footnotesize  $E_7 \times SU(2)$}    & $ \frac12$  \\
7 &\small SO(10) GUT &\small $(3,2,1, 5,1)$ & $30$ & {\footnotesize $E_8$} &  1 \\
 \hline
\end{tabular}
\vskip .7truecm
\caption{\small List of all Standard Model extensions and the resulting
CFT charge quantization (last column).  In the fifth column we display the gauge group of the lifted theory, using the
first lift ({\it i.e.} $SU(5)\times SO(10)$). We
have omitted the $SO(10)$ factor, which is never combined with other parts of the gauge group; we have also omitted $U(1)$ factors
originating from the $N=2$ minimal models, or sporadically occurring extensions of those factors.}
\label{TableG}
\end{center}
\end{table}
}

\renewcommand{\arraystretch}{1.2}
{\small
\begin{table}[h]
\begin{center}
\vskip .7truecm
\begin{tabular}{|c|c|c||c|c|c|c|}
\hline
\hline
Nr. & Unlifted & Current & Order & Gauge group    & Charge quantum \\
\hline
0 &\small SM, Q=1/6& \small $(1,1,0,1,1)$ & $1$ & {\footnotesize  $SU(3)\times SU(2) \times U(1) \times SO(16)\times U_{20}$} &    $ \frac16$  \\
1 &\small SM, Q=1/3 &\small $(1,2,15,1,1)$ & $2$ &  {\footnotesize $SU(3)\times SU(2)\times U(1)\times SO(16)\times U_{20}$}  &    $ \frac13$  \\
2 &\small SM, Q=1/2 &\small $(3,1,10,1,1)$& $3$ &  {\footnotesize $SU(3)\times SU(2)\times U(1)\times SO(16)\times U_{20}$}  &    $ \frac12$  \\
3 &\small LR, Q=1/6 &\small $(1,1,6,1,4)$ & $5$ & {\footnotesize $SU(3)\times SU(2) \times SO(16)\times SU(2)\times U_{12}$}     & $ \frac16$  \\
4 &\small SU(5) GUT&\small $(\bar 3,2,5,1,1)$ & $6$ & {\footnotesize  $SU(5) \times SO(16)\times U_{20}$}     & 1  \\
5 &\small LR, Q=1/3 &\small $(1,2,3,{16},2)$ & $10$ &  {\footnotesize $SU(3)  \times SO(20) \times U_{12}$     }  &    $ \frac13$  \\
6 &\small Pati-Salam &\small $(\bar 3,0,2,{16},-2)$ & $15$ & {\footnotesize  $SO(22)\times SU(2)\times SU(2)$}    & $ \frac12$  \\
7 &\small SO(10) GUT &\small $(3,2,1,1,4)$ & $30$ & {\footnotesize $SO(26)$} &  1 \\
 \hline
\end{tabular}
\vskip .7truecm
\caption{\small Same as Table \ref{TableG}, but for the second lift, $SO(16)\times U(1)_{20}$, corresponding to the last two entries in column 3. Note that in these two factors 
there is an extension by the current $(16,10)$. This has to be taken into account in order to compute the currents as powers of the generator of the orbit.}
\label{TableG2}
\end{center}
\end{table}
}

Note that, by construction, $SU(3)_{\rm color}$ always has to come from the first factor in the gauge group listed in column 5. There may well exist other
interesting ways to embed $SU(3)\times SU(2)\times U(1)_Y$ into $SO(10)\times E_8$, but we will only consider here the one we get
using B-L lifting. In cases 6 and 7 the
gauge group has no chiral representations. In cases 3 and 5 it does have chiral representations, but none that can yield a standard model
family. In case 5 the $E_8$ group breaks in the familiar way to $SU(3)\times E_6$, but unlike standard Calabi-Yau  
compactifications the standard model color group is now embedded in $SU(3)$, whereas $SU(2)\times U(1)$ is embedded in $E_6$. 
This implies immediately that the usual correlation between color and $Y$-charge is lost. Indeed, the representations $(3,1)$ and 
$(1,27)$ contain fractionally charged matter, and the same is true for the combination $(3,27)$. Hence there is no way to get
chiral standard model families out of this embedding.  

In case 3, the representation $(6)$ of $SU(6)$ provides $Y$-charges $\frac16$ and $-\frac56$, the $(15)$ gives charges
$\frac23$ and $-\frac43$ and the anomaly-free representation $(20)$ breaks to the real combination $(3,\overline{10})+(-3,10)$ of
$U(1)_Y \times SU(5)$. Hence only the  $(20)$ can provide the charges of the weak lepton doublet, but since it is real this is not possible.
Therefore this possibility can also be eliminated from the list. This leaves precisely all the cases where the extra
$SU(5)$ factor obtained from lifting $B-L$ (denoted $SU(5)_{\rm B-L}$ henceforth) does not mix with the standard model gauge group. 
The remaining possibilities, the only ones that can
yield chiral families, are
$SU(5)$ GUTs and three versions of the standard model with electric charge quantization in units of $\frac16$, $\frac13$ and $\frac12$.

The $SU(5)$ models (Nr. 4 in Table \ref{TableG}) come without massless scalars  in the $(24)$ representation that could act as Higgses to break the GUT group
to the standard model. One could still postulate that $(24)$'s are created dynamically as bound states
of bi-fundamentals with one component in the $SU(5)_{\rm GUT}$ and another in either $SU(5)_{\rm B-L}$ or $SO(10)$, but this is
a far-fetched scenario.  Therefore these $SU(5)_{\rm GUT}$ models don't look very promising. On the other hand, the other three, nrs. 0, 1 and 2,
have an unextended gauge group $SU(3)\times SU(2) \times U(1)_Y$ multiplied with a ``hidden" gauge group. The latter is
$SU(5)\times SO(10) \times G_{\rm int}$, where $G_{\rm int}$ is generated by spin-1 currents originating from the minimal tensor
factors. Experience with the standard Gepner models and their Calabi-Yau analogs suggests that this latter gauge group is usually
broken by moduli, since it is not present in generic points in the moduli space of Calabi-Yau manifolds.

The models that cannot produce the standard model spectrum in B-L-lifted Gepner models, nrs. 3, 5, 6 and 7 in Table \ref{TableG}, 
correspond to the class of ``left-right" symmetric models in standard Gepner models;
in addition to the weak gauge group they have a group $SU(2)_R$. This includes the Pati-Salam model. In \cite{GatoRivera:2010gv} we
explained how the $SU(2)_R$ extension could be avoided. It requires the current $(1,1,6,4)$ which combines $U(1)_Y$ and
$U(1)_{B-L}$ to be mapped to a different current in the bosonic sector. Since $(1,1,6,4)$ has order 5 we must look for an order
5 current in the bosonic sector. The canonical choice  ($(1,1,6,4)$ in standard Gepner models or $(1,1,6,5,1)$ in B-L lifted ones)
will not do, and neither will its conjugate. These simply extend the gauge group to $SU(2)_R \times U(1)$, respectively $SU(6)$. So what needs
to be done is to look for a current with components inside the internal Gepner model sector. Such a current can only exist if at least
one of the Gepner model factors has a $Z_5$ simple current symmetry (in fact, then  there will always be at least two such factors, 
since otherwise their central charges cannot add up to nine). This condition is satisfied by 66 of the 168 Gepner combinations,
plus 20 of the 59 Gepner combinations with an additional exceptional $SU(2)$ invariant. This has also implications for the possibility
of successfully applying B-L lifting in other classes of heterotic strings. 
Note, for example, that although B-L lifting is applicable
to free fermionic constructions, this will not lead to chiral spectra, because there are no order five currents available.

 A table analogous to Table \ref{TableG} can be made for the second non-trivial B-L lift and is shown in Table \ref{TableG2}.
 Although the gauge groups are different, the
 conclusion is essentially the same with regard to the standard model spectra: only nrs. 0, 1, 2 and 4 can occur with chiral
 families. In all other cases the full gauge group either has no chiral representations at all, or the standard model is
 embedded in such a way that there is no way to get a standard family. The least obvious case is nr. 3. Here
 the standard model $U(1)$ factor $Y$ is embedded in $SO(16)\times SU(2) \times U_{12}$, extended by the current
 $(v,1,6)$, which originates from the extension current $(v,10)$ of the lift CFT $SO(16)\times U_{20}$. In a standard left-right
 symmetric model of type SM, $Q=\frac16$, $Y$ is embedded in $SU(2) \times U_{12}$, and all standard model charges 
 can be obtained for suitable representations of that group. But if this algebra is multiplied with $SO(16)$ and extended by
 $(v,1,6)$, some of the necessary $SU(2) \times U_{12}$ are projected out, because they have half-integer charge with
 respect to the current $6$ of $U_{12}$. Then they can only appear in combination with the $SO(16)$ spinor, so that the total conformal weight
exceeds one. This happens, for example, for $Y$-charge 6, needed for the charged lepton singlets. Hence a massless standard model representation
cannot be obtained in this case.

 \section{Results}
 
 In Table \ref{BLSummary} we show the results 
 for all 86 combinations for which chiral standard model spectra are
 in principle possible. These results are based on spectra obtained using a randomly chosen simple current group with
 at most four generators, and a randomly chosen discrete torsion matrix, as defined in \cite{Kreuzer:1993tf}. 
 In our scan the standard model gauge group is input, but the standard model representations are not. The former is implied by
 the affine Lie algebras we use to build the bosonic sector of the heterotic string, which contains $SU(3)\times SU(2) \times U(1)$.
The chiral representations we get can be just  a number of chiral families, but can also contain vector-like
or chiral fractionally charged states  \cite{Wen:1985qj,Athanasiu:1988uj}
which must necessarily be
present in the spectrum of these models \cite{Schellekens:1989qb}, but are not necessarily massless. 
 As in our previous work, \cite{GatoRivera:2010gv}
and  \cite{GatoRivera:2010xn}, we first eliminate all spectra with {\it chiral} fractionally charged particles. 
In those cases, there is
not even a well-defined notion of the number of chiral families. For the remaining spectra, in column 2 we list the greatest
common divisor $\Delta$ of the number of families, in column 3 the maximum observed number of families, in column 4
 the number of distinct MIPFs with three families, and in column 5 the number of distinct MIPFs with $N$ families (including $N=3$ but excluding $N=0$).  The MIPFs
are distinguished on the basis of certain features of the spectrum, in the same way as in \cite{GatoRivera:2010gv}
and  \cite{GatoRivera:2010xn}.  They are the group type (as defined in Table \ref{TableG}), the number of 
vector-like pairs of Q, U, D, L, E quarks and leptons, the total number of singlets,
the total size of the gauge group,  the total amount of {\it vector-like} fractionally charged matter and the fractional charge quantum, in as far
as it is not the one determined by the group type. Indeed, in rare cases the charge quantum of the massless spectrum is larger than 
the one listed in the last column of Table \ref{TableG}, which implies that the smaller quanta only occur in the massive spectrum. The case of
most interest is the one where all massless matter is integrally charged (for color singlets) without GUT unification (which, as explained above, cannot be broken).
In \cite{GatoRivera:2010xn} we found examples of this kind for an even number of families; in the present work we only have
examples with two families. Examples with three families were reported recently in  \cite{Assel:2009xa}. 

To get an idea of the amount of saturation of our random scan we have checked the completeness of the list under mirror
symmetry, which is guaranteed to be an exact symmetry of the complete list. Therefore the number of missing mirror spectra
will go to zero as the list approaches completeness. The percentage of $N$-family spectra with missing mirrors is
shown in the last column of Table \ref{BLSummary}.  The numbers in columns 4 and 5 are the number of families after
identifying mirror spectra (so that each mirror pair, complete or incomplete, is counted just once). 

We have done a similar scan for the second B-L-lift, the one that corresponds to an embedding of the standard model
in $SO(32)$. The result is that chiral standard model spectra occur in precisely the same tensor combinations as above, but it turns
out that $\Delta$ is a multiple of either 2 or (in a few cases) 4 or 8. Hence three families cannot be obtained with this lift.

\LTcapwidth=14truecm
\begin{center}
\vskip .7truecm
\begin{longtable}{|c||c|c|c|c|c|}\caption{{\bf{Results for Gepner models with B-L lifting}}}\\
\hline
 \multicolumn{1}{|c||}{model}
& \multicolumn{1}{c|}{$\Delta$}
& \multicolumn{1}{l|}{Max. }
& \multicolumn{1}{c|}{3 family}
& \multicolumn{1}{c|}{$N$ fam. ($N>0$)}  
& \multicolumn{1}{c|}{Missing Mirrors } \\ 
\hline
\endfirsthead
\multicolumn{6}{c}%
{{\bfseries \tablename\ \thetable{} {\rm-- continued from previous page}}} \\
\hline 
 \multicolumn{1}{|c||}{model}
& \multicolumn{1}{c|}{$\Delta$}
& \multicolumn{1}{l|}{Max. }
& \multicolumn{1}{c|}{3 family}
& \multicolumn{1}{c|}{$N$ fam. ($N>0$)}  
& \multicolumn{1}{c|}{Missing Mirrors } \\ 
\hline
\endhead
\hline \multicolumn{6}{|r|}{{Continued on next page}} \\ \hline
\endfoot
\hline \hline
\endlastfoot\hline
\label{BLSummary}
$(1,10,13,58)$ & 1 & 20 & 69 & 492 &   19.92\% \\ 
$(1,10,18,28)$ & 1 & 11 & 97 & 536 &   10.26\% \\ 
$(1,1,1,1,8,13)$ & 2 & 10 & 0 & 16 &    6.25\% \\ 
$(1,1,1,2,3,18)$ & 0 & 0 & 0 & 0 &     -\\ 
$(1,1,1,3,3,8)$ & 0 & 0 & 0 & 0 &     -\\ 
$(1,12,13,33)$ & 0 & 0 & 0 & 0 &     -\\ 
$(1,1,2,13,58)$ & 1 & 12 & 16 & 124 &   12.10\% \\ 
$(1,1,2,18,28)$ & 1 & 8 & 19 & 220 &   13.64\% \\ 
$(1,1,3,10,18)$ & 0 & 0 & 0 & 0 &     -\\ 
$(1,1,3,13,13)$ & 1 & 30 & 54 & 414 &   16.91\% \\ 
$(1,13,13,28)$ & 1 & 24 & 147 & 1021 &   18.12\% \\ 
$(1,13,18,18)$ & 1 & 24 & 29 & 256 &    0 \\ 
$(1,1,3,6,118)$ & 0 & 0 & 0 & 0 &     -\\ 
$(1,1,3,7,43)$ & 1 & 15 & 1 & 19 &   21.05\% \\ 
$(1,1,3,8,28)$ & 1 & 24 & 95 & 764 &    8.64\% \\ 
$(1,1,4,8,13)$ & 1 & 10 & 6 & 62 &    8.06\% \\ 
$(1,2,2,8,13)$ & 0 & 0 & 0 & 0 &     -\\ 
$(1,2,3,3,58)$ & 1 & 12 & 100 & 433 &   13.86\% \\ 
$(1,2,3,4,18)$ & 2 & 12 & 0 & 77 &    0 \\ 
$(1,3,3,3,13)$ & 1 & 21 & 146 & 1268 &   11.44\% \\ 
$(1,3,3,4,8)$ & 1 & 12 & 36 & 177 &   12.43\% \\ 
$(1,5,43,628)$ & 1 & 14 & 0 & 21 &    0 \\ 
$(1,5,58,138)$ & 1 & 10 & 0 & 26 &    0 \\ 
$(1,5,68,103)$ & 1 & 10 & 2 & 16 &    0 \\ 
$(1,6,23,598)$ & 1 & 8 & 4 & 45 &    4.44\% \\ 
$(1,6,28,118)$ & 1 & 12 & 102 & 861 &   26.02\% \\ 
$(1,6,38,58)$ & 1 & 11 & 30 & 221 &   12.67\% \\ 
$(1,7,18,178)$ & 1 & 7 & 15 & 92 &    0 \\ 
$(1,7,28,43)$ & 1 & 8 & 18 & 117 &    1.71\% \\ 
$(1,8,14,238)$ & 1 & 14 & 25 & 263 &   22.43\% \\ 
$(1,8,16,88)$ & 1 & 15 & 66 & 593 &   14.50\% \\ 
$(1,8,18,58)$ & 1 & 24 & 600 & 4004 &   17.63\% \\ 
$(1,8,22,38)$ & 1 & 8 & 5 & 129 &    4.65\% \\ 
$(1,8,28,28)$ & 1 & 36 & 1284 & 7355 &   24.00\% \\ 
$(1,9,13,108)$ & 0 & 0 & 0 & 0 &     -\\ 
$(2,2,2,3,18)$ & 1 & 18 & 7 & 280 &    7.50\% \\ 
$(2,2,3,3,8)$ & 1 & 16 & 81 & 587 &    2.21\% \\ 
$(2,3,19,418)$ & 0 & 0 & 0 & 0 &     -\\ 
$(2,3,20,218)$ & 2 & 30 & 0 & 72 &    5.56\% \\ 
$(2,3,22,118)$ & 1 & 33 & 35 & 182 &   10.99\% \\ 
$(2,3,23,98)$ & 1 & 33 & 58 & 453 &    0.88\% \\ 
$(2,3,26,68)$ & 1 & 24 & 3 & 83 &    4.82\% \\ 
$(2,3,28,58)$ & 1 & 48 & 468 & 3104 &   14.43\% \\ 
$(2,3,34,43)$ & 0 & 0 & 0 & 0 &     -\\ 
$(2,3,38,38)$ & 1 & 66 & 199 & 1911 &   14.08\% \\ 
$(2,4,13,58)$ & 1 & 10 & 38 & 202 &    6.44\% \\ 
$(2,4,18,28)$ & 1 & 9 & 16 & 186 &    3.23\% \\ 
$(2,5,8,138)$ & 1 & 10 & 4 & 37 &    8.11\% \\ 
$(2,6,8,38)$ & 1 & 10 & 32 & 329 &    2.74\% \\ 
$(2,8,10,13)$ & 2 & 8 & 0 & 34 &    0 \\ 
$(2,8,8,18)$ & 1 & 18 & 375 & 2481 &    1.89\% \\ 
$(3,3,10,58)$ & 1 & 18 & 33 & 274 &    0 \\ 
$(3,3,12,33)$ & 1 & 16 & 0 & 38 &    0 \\ 
$(3,3,13,28)$ & 1 & 33 & 683 & 3871 &    5.11\% \\ 
$(3,3,18,18)$ & 1 & 42 & 1379 & 8193 &    3.23\% \\ 
$(3,3,3,3,3)$ & 1 & 30 & 1725 & 11574 &   10.73\% \\ 
$(3,3,9,108)$ & 1 & 20 & 4 & 52 &    1.92\% \\ 
$(3,4,10,18)$ & 2 & 12 & 0 & 78 &    0 \\ 
$(3,4,13,13)$ & 1 & 30 & 23 & 143 &    4.20\% \\ 
$(3,4,6,118)$ & 1 & 14 & 21 & 200 &    2.50\% \\ 
$(3,4,7,43)$ & 1 & 15 & 0 & 6 &    0 \\ 
$(3,4,8,28)$ & 1 & 27 & 386 & 2450 &    5.18\% \\ 
$(3,5,5,68)$ & 1 & 12 & 0 & 20 &    0 \\ 
$(3,6,6,18)$ & 1 & 18 & 2 & 159 &    3.77\% \\ 
$(3,8,8,8)$ & 1 & 45 & 3756 & 22618 &    5.73\% \\ 
$(4,4,8,13)$ & 2 & 12 & 0 & 68 &    0 \\ 
$(1,10,18,28^*)$ & 0 & 0 & 0 & 0 &     -\\ 
$(1,1,2,18,28^*)$ & 0 & 0 & 0 & 0 &     -\\ 
$(1,1,3,10^*,18)$ & 0 & 0 & 0 & 0 &     -\\ 
$(1,13,13,28^*)$ & 1 & 18 & 35 & 299 &   10.03\% \\ 
$(1,1,3,8,28^*)$ & 0 & 0 & 0 & 0 &     -\\ 
$(1,6,28^*,118)$ & 0 & 0 & 0 & 0 &     -\\ 
$(1,7,28^*,43)$ & 1 & 9 & 1 & 13 &    7.69\% \\ 
$(1,8,16^*,88)$ & 1 & 10 & 7 & 68 &   17.65\% \\ 
$(1,8,28,28^*)$ & 1 & 14 & 35 & 318 &   10.06\% \\ 
$(2,3,28^*,58)$ & 1 & 12 & 53 & 185 &   14.05\% \\ 
$(2,4,18,28^*)$ & 2 & 8 & 0 & 11 &    0 \\ 
$(2,8,10^*,13)$ & 0 & 0 & 0 & 0 &     -\\ 
$(3,3,10^*,58)$ & 1 & 12 & 26 & 114 &    0 \\ 
$(3,3,13,28^*)$ & 1 & 21 & 124 & 1045 &   11.10\% \\ 
$(3,4,10^*,18)$ & 2 & 12 & 0 & 21 &    0 \\ 
$(3,4,8,28^*)$ & 1 & 12 & 23 & 99 &    3.03\% \\ 
$(1,10^*,18,28)$ & 1 & 5 & 8 & 61 &    0 \\ 
$(1,8,28^*,28^*)$ & 0 & 0 & 0 & 0 &     -\\ 
$(1,10^*,18,28^*)$ & 0 & 0 & 0 & 0 &     -\\ 
\end{longtable}
\end{center}

\begin{figure}
\begin{center}
\includegraphics[width=17cm]{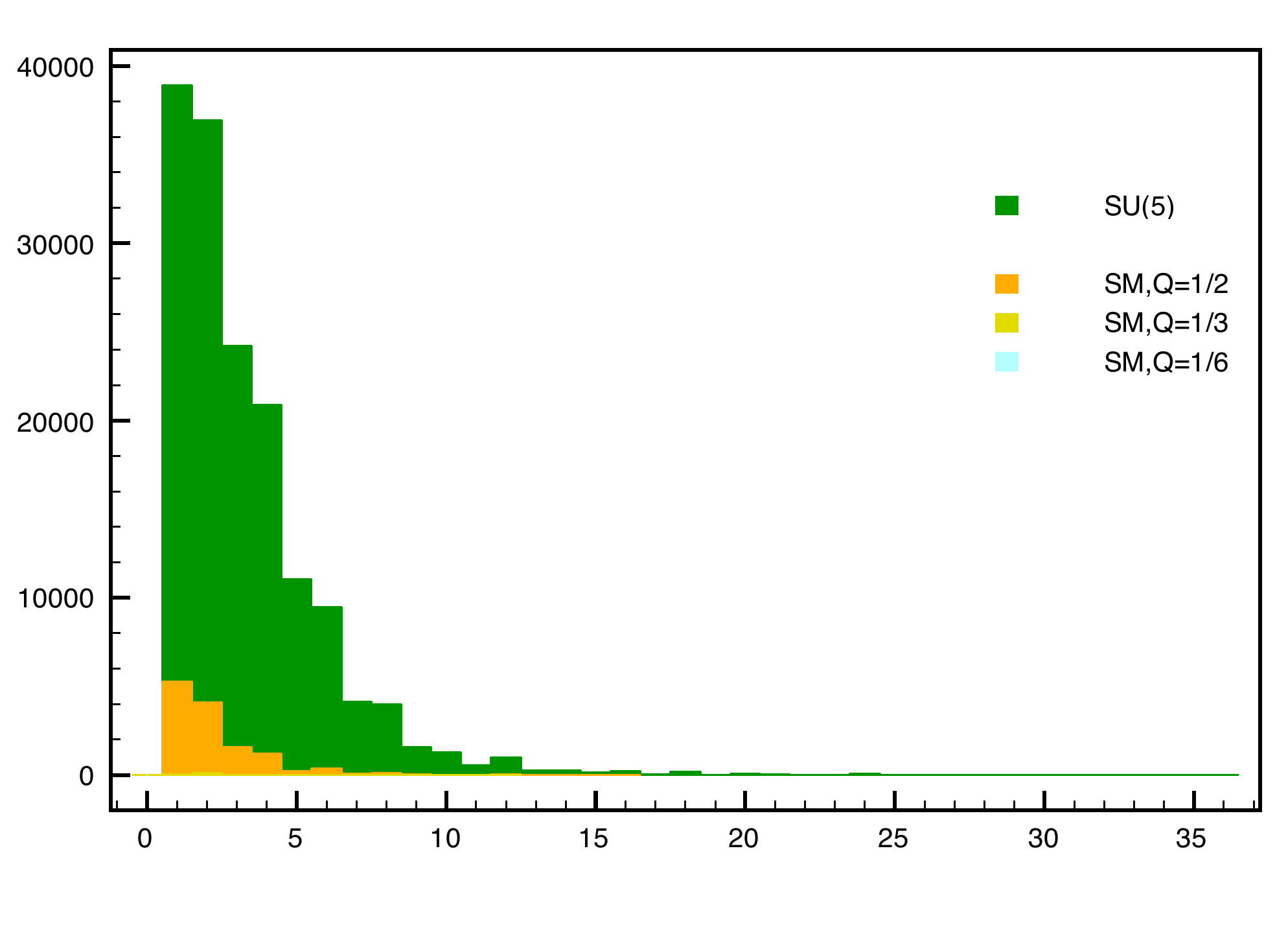}
\caption{Distribution of the number of families for all tensor products combined.}
\label{famplot}
\end{center}
\end{figure}

In figure \ref{famplot} we show the family distribution for the first lift, based on family numbers without
mirror identification, the same quantity that was plotted in \cite{GatoRivera:2010xn} for heterotic weight
lifting. We also use the same color coding for group types, so that this plot can be directly compared
to the corresponding one in \cite{GatoRivera:2010xn}. All these plots are stacked histograms, {\it i.e.} the bars of different
colors are stacked on top of each other, and cannot be hidden behind others.
Apart from the fact that some group types and colors 
are absent in the present case, these plots look very similar. In particular, three families occur as part of a 
fairly normal distribution, only slightly suppressed with respect to the even family numbers. This conclusion
is weakened if we were to include also the second lift, which only produces even family numbers. However, if we include
both lifts with equal weight, this is still nowhere near the two to three orders of magnitude suppression observed in orientifold models 
\cite{Dijkstra:2004cc,Gmeiner:2005vz}.

In total, we scanned $75 \times 10^6$ spectra using the first B-L lift. Of these, 
$61 \times 10^6$ are
free of chiral exotics, and $23 \times 10^6$ have a non-zero number of families. However, most of these are $SU(5)$ GUT models.
The chiral spectra without chiral exotics are distributed as follows over the various group types: $98.87\%$ $SU(5)$, 
$1.11\%$ SM, Q=$\frac12$ and $0.02\%$ SM,Q=$\frac13$. We found none with SM,Q=$\frac16$. For comparison, for heterotic
weight lifting these percentages are respectively $98.9\%$, $0.95\%$, $0.069\%$ and $0.0001\%$, after removing the four group
types that cannot occur for B-L lifting, and normalizing the sum to $100\%$. In the total sample, about $0.3\%$ of the spectra has
no chiral exotics, chiral families and no $SU(5)$ GUT unification.
As in \cite{GatoRivera:2010xn}, all these percentages are based on total occurrence
counting, and not on counting distinct MIPFs. 

\begin{figure}
\begin{center}
\includegraphics[width=17cm]{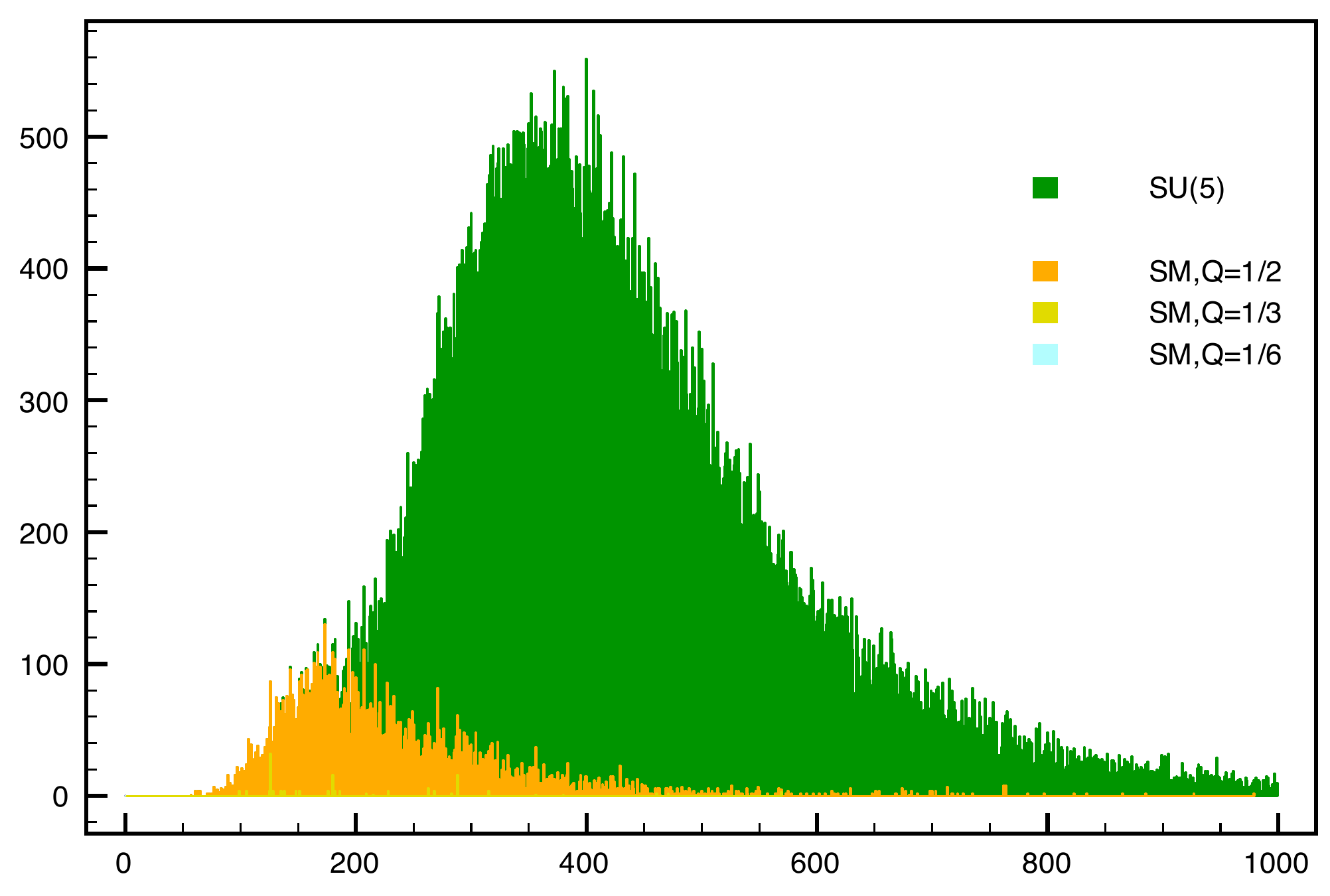}
\caption{Distribution of the number of standard model singlets}
\label{sing}
\end{center}
\end{figure}

\begin{figure}
\label{absing}
\begin{center}
\includegraphics[width=17cm]{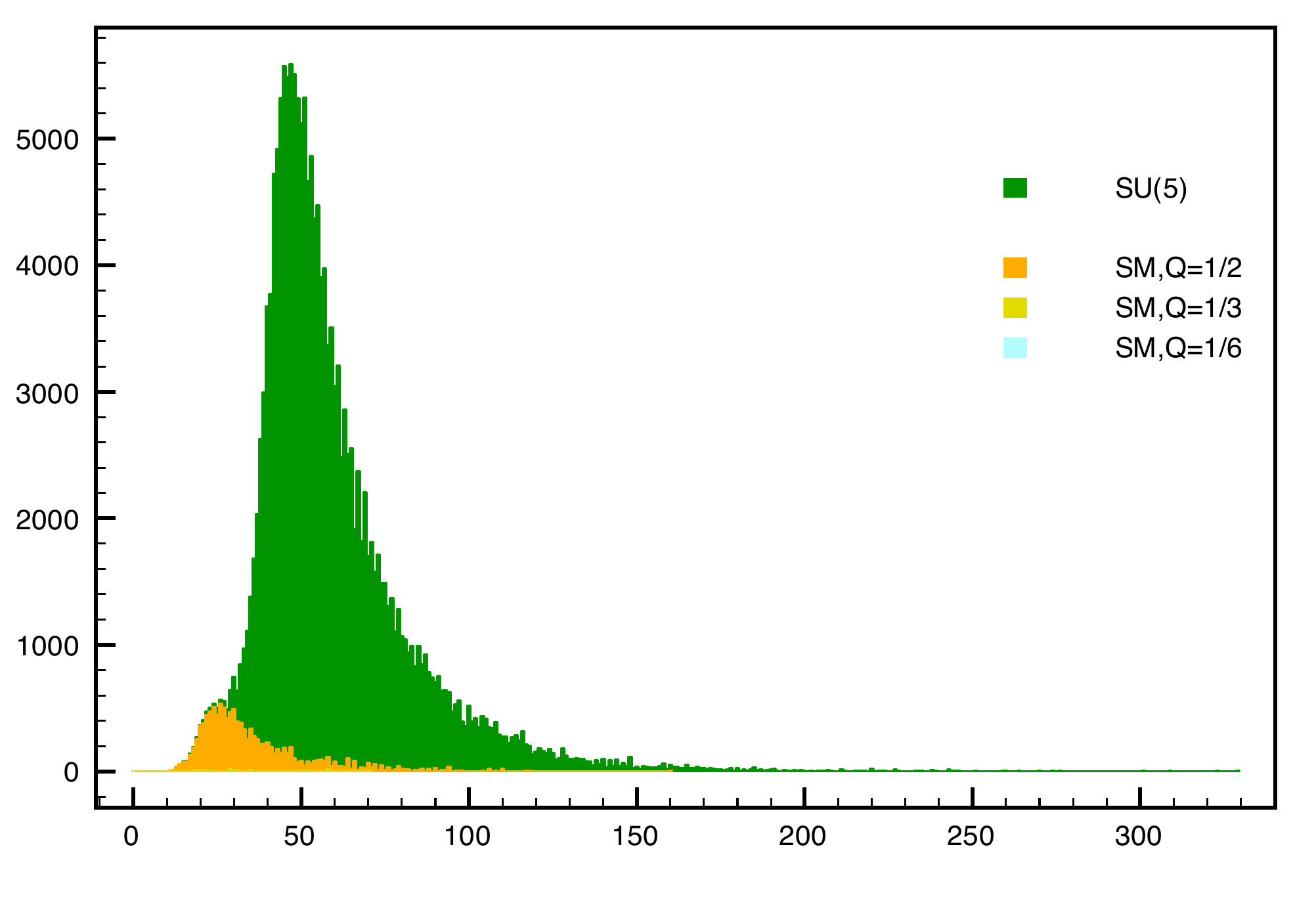}
\caption{Distribution of the number of standard model singlets that do not couple to any non-abelian hidden sector group}
\end{center}
\end{figure}

None of these results are significantly different for B-L lifting and heterotic
weight lifting, and the same is true for most of the other distributions as well. 
There is one exception 
concerning the coupling of matter to additional non-abelian gauge group factors. In the present case, 
these non-abelian gauge group factors are the $SU(5)\times SO(10)$ groups originating from the lift. We will not consider
sporadically occurring  extensions of the chiral algebra that may enlarge the $N=2$ minimal model $U(1)$ factors to a non-abelian gauge group.
 In \cite{GatoRivera:2010xn}
a plot was shown for the distribution of the number of ``abelian singlets": standard model singlets that are also singlets with respect to all
non-abelian groups (these particles can be charged under the abelian factors in the gauge group, namely B-L and the
minimal models $U(1)$'s).
This plot had an interesting (and not yet understood) double peak structure, and a significant
tail towards zero. In particular, there was a substantial number of examples without any such singlets, in other 
words examples where all matter couples to at least one non-abelian group, or is charged under $U(1)_Y$. This does
not happen in B-L lifted models. The distribution for standard model singlets (regardless of any other interactions) is similar
to those in  \cite{GatoRivera:2010gv} and \cite{GatoRivera:2010xn} and is shown in figure \ref{sing}. The distribution for abelian
singlets,   shown in figure \ref{absing}, has the same shape, shifted towards zero, and has no double peak or a tail reaching zero. The latter feature is probably a
consequence of the fact that B-L lifting, unlike heterotic weight lifting, cannot affect the internal CFT contribution to the conformal weight  (the part coming from 
the internal $c=9$ CFT). Therefore standard model singlets that have vanishing B-L charge will remain singlets, and there are always
some particles of that kind. Since the first B-L lift has a completely non-abelian gauge group, 
any representation that has non-vanishing B-L charge will be lifted to a non-trivial
$SU(5)\times SO(10)$ representation (see Table \ref{TableLift}).  A similar phenomenon can be seen in the distribution of ``abelian
fractionals", fractionally charged particles that are singlets under all non-abelian groups outside the standard model, and hence
cannot possibly be confined. In the case of heterotic weight lifting this plot had a large peak at zero for spectra with half-integer
charge quantization. In the present case the distribution looks similar, but the peak at zero is absent.

An important question is whether the standard model families couple to the non-abelian factors $SU(5)_{\rm B-L}$ or $SO(10)$ in 
the hidden sector, in which case ``hidden" would be a misnomer. Since we presented these factors as the lift of $U(1)_{\rm B-L}$, and
since in standard $SO(10)$ models the quarks and leptons do indeed carry a non-zero $B-L$ charge, one might worry that this
charge gets lifted to a non-trivial $SU(5)_{\rm B-L} \times SO(10)$ representation. However, this cannot happen for a three-family
model without mirror fermions, because the smallest non-trivial representations of $SU(5)_{\rm B-L} \times SO(10)$ have dimension five.
Therefore the only way standard model matter could couple to $SU(5)_{\rm B-L} \times SO(10)$ is if for a given quark or lepton there
are at least two mirror pairs. The plot for the percentage of matter that does not couple to the non-abelian part of the hidden sector looks
very similar to the one presented in \cite{GatoRivera:2010xn}, and in particular has a huge peak at $100\%$. Therefore there are many examples
without such particles, and we will present one in the next chapter.

\section{Examples}



%




We found a total of 12620 distinct three-family spectra (24203 prior to identification of mirrors) with a gauge group $SU(3)\times SU(2)\times U(1)$ with
charge quantization in half-integer units. Apart from the standard model group, the gauge group
of these models has the usual factors $SU(5)_{B-L}$, $SO(10)_{B-L}$ and $U(1)^n$, where $n$ is the number
of  $N=2$ minimal models in the tensor product. Occasionally there may be further extensions. 

One of the simplest
examples is the following one, obtained for the tensor product $(3,3,3,3,3)$. We denote the standard model representations
for left-handed quarks and leptons as $Q, U^c, D^c, L$ and  $E^c$, where $Q=(3,2,\frac16)$, etc. The last two entries are
respectively the $SU(5)_{B-L}$ and $SO(10)_{B-L}$ representations. 
In this example the quarks, leptons, and mirrors are
\begin{eqnarray*}
 &3 \times  (Q,1,1) + 3 \times(U^c,1,1)+ 5 \times (D^c,1,1)+2 \times (D,1,1) \\ 
 &+ 5 \times (L,1,1)+ 2 \times (L^c,1,1) + 3 \times(E^c,1,1)
 \end{eqnarray*}
 Note that they do not couple to the non-abelian hidden groups. There are two supersymmetric Higgs pair candidates $L+L^c$,
 and two vector-like $D+D^c$ pairs, but apart from this the quark and lepton spectrum is exactly that of the standard model.
 The standard model singlets $N$ are 
 \begin{eqnarray*}
\\ &24  \times  (N,1,1)
+ 6   \times  (N,5,1) 
+ 7   \times  (N,\bar{5},1) 
+  (N,10,1) 
+ 5 \times (N,1,10)
+  (N,1,\overline{16}) 
 \end{eqnarray*}
 Note that the $SU(5)_{B-L}$ representation is $7 \times (\bar 5)+ 6 \times (5) + (10)$, which is anomaly free, and corresponds to
 one net family of  $SU(5)_{B-L}$.
 The fractionally charged matter was required to be vector-like with respect to $SU(3)\times SU(2)\times U(1)$, but
 not
 with respect to the rest of the gauge group. Indeed, in this example we find that it is chiral with respect to $SO(10)_{B-L}$. 
 The part that is  vector-like with respect to $SO(10)_{B-L}$ is
  \begin{eqnarray*}
&\left[  
2 \times           (1,1,-\frac12,5,1)
+   3 \times           (1,1,\frac12,5,1) 
+  15 \times           (1,1,\frac12,1,1)
+    2 \times           (3,0,\frac16,1,1) + {\rm c.c} \ \right])
\\
&
+\ 4  \times           (1,2,0,1,1)
 \end{eqnarray*}
In addition, there are the following representations that are chiral with respect to $SO(10)_{B-L}$
 \begin{eqnarray*}
(1,1,\frac12,1,\overline{16})+(1,1,-\frac12,1,\overline{16})
 \end{eqnarray*}

 We see  that in the ``hidden" sector there is one net family of $SU(5)_{B-L}$ and three (!) of $SO(10)_{B-L}$ (namely a total of three $(\overline{16})$'s). 
 This group, $SO(10)_{B-L}$, is in fact
 the gauge group we would use to embed the standard model in standard Gepner constructions. It may seem strange that
 we get three families from the ``quintic" (the combination (3,3,3,3,3)), which is known to give only an even number of families in standard Gepner constructions. 
 The reason is very simple. In this example the $E_8$ factor is broken to a subgroup. In a subsequent paper, \cite{GatoRivera:EBR}, we
 will investigate this possibility in more detail. It turns out that just breaking $E_8$ to $SU(5) \times SU(5)$ in combination
 with the quintic gives rise to a large number of three family models. 
 
 It is clear that the chiral $SO(10)_{B-L}$ symmetry forbids a mass for the last two fractionally charged particles. However, if one of
 the five scalars $(N,1,10)$ gets a vacuum expectation value this will break $SO(10)$ to a harmless $SO(9)$ symmetry. In addition to
 $SO(10)$ also the $N=2$ abelian gauge symmetries $U(1)^5$ will have to be broken in order to give a mass to the
 fractionally charged matter. For example, the $U(1)$ charges of the last two representations are
 $$ (1,1,\frac12,1,\overline{16},0,0,\frac{1}{10},\frac{3}{10},-\frac{3}{10})+(1,1,-\frac12,1,\overline{16},-\frac{3}{10},-\frac{3}{10},-\frac{1}{10},0,0) \ ,$$
 where the last five entries indicate the charges. As long as these five $U(1)$'s are unbroken, charge conservation
 forbids most low order couplings, including masses for particles that are vector-like with respect to the standard model. 
 The Yukawa couplings in the exact Gepner point are also restricted by the $U(1)^n$ charges
 of the quarks, leptons and the Higgs.  However, if we allow vacuum expectation values for the large number of 
 standard model singlets, all of which in general carry different sets of charges, there is in principle no obstacle to generate
 missing couplings from higher order couplings saturated with scalar vevs. It is clear that most of this breaking will have to
 take place near the Planck or string scale. Otherwise most relevant couplings, and in particular the masses of the
 fractionally charged particles, will simply be too small. In other words, the exact RCFT gives a very special point in
 moduli space, and it is extremely unlikely that this would be precisely what we need. 
 
%
Note that these spectra contain matter that couples to the standard model gauge bosons as well as to the
$SU(5)$ and $SO(10)$ ``hidden sector" gauge bosons. Furthermore there is matter of this kind that is chiral with respect to the hidden sector.
This sort of matter was not allowed in the orientifold model searches of  \cite{Dijkstra:2004cc}  and   \cite{adks}, but if it had been allowed
spectra containing matter of this kind would have dominated the set of solutions by a large factor. In the present work, we have not searched
systematically for spectra with only vector-like ``observable-hidden matter" (in orientifold language), because we have tracked only
the total dimensions of  the $SU(5)$ and
$SO(10)$ representations. Only in explicit examples, such as this one, the representations themselves can be seen. Since in $SU(5)$ and
$SO(10)$ models the zero-family case dominates all others, we expect examples to exist where all  particles are vector-like with respect to
$SU(5)_{B-L}$ as well as $SO(10)_{B-L}$, but without further study we cannot say how frequently they occur.  
In the present example, there are also some standard model singlets that couple chirally to the hidden sector.
This is a much less serious problem, and such particles were allowed in the aforementioned orientifold model searches. 
 
 A problem one may have to worry about in B-L lifted models is the absence of $U(1)_{B-L}$. It is replaced by non-abelian groups
 that do not couple to quarks and leptons.
 By itself, the absence of a massless gauge boson coupling to $B-L$ is of course a good 
 feature.  However, a discrete subgroup of $U(1)_{B-L}$ is often used in low energy supersymmetry scenarios as part of an R-parity symmetry
 that forbids highly dangerous baryon and/or lepton  number violating dimension four operators. Although we have not committed ourselves here
 to  a particular choice of the supersymmetry breaking scale, in heterotic models light squarks, sleptons and gauginos are
 probably needed to make the gauge couplings reach the correct low-energy values. 
 In the example above there are only a few other
 candidates for matter that could live in the GUT desert and hence could influence the running of couplings: two $(D,D^c)$ mirror pairs and an additional Higgs.
 The fractionally charged matter cannot contribute significantly because
 the lightest fractionally charged particle  would be stable, and almost certainly too abundant, if it had a mass below the GUT scale. 
 However, even if the correct running of couplings from the unification point (which anyhow is two orders
  of magnitude too large in heterotic strings)
  requires low-energy supersymmetry, which in its turn might imply undesirable couplings, it is still possible that remnants of the $U(1)^5$ symmetries provide the required
  selection rules to forbid those couplings.     Of course this can all be checked in more detail, but the particular model shown here is just one of many, selected only on the
  basis of having relatively few vector-like states and no couplings of the quarks and leptons to the hidden sector gauge bosons. 
  

The foregoing example is one out of the 1560 distinct three family models of type SM, Q=$\frac12$. The total number of distinct spectra of this
type with $N$ families ($N > 1$ and including 3) is 12895 (these numbers are prior to mirror identification). Hence three families are found in about $12\%$ of these cases.
The number of spectra of type  SM, Q=$\frac13$ is so small that it is barely visible in the plots. In total, we found 140 spectra of this type,
and one one of them has three families. It occurs for the tensor product $(2,3,28,58)$ and it has the following spectrum.

Quarks, leptons, and mirrors:
\begin{eqnarray*}
 &3 \times  (Q,1,1) + 3 \times(U^c,1,1)+ 8 \times (D^c,1,1)+5 \times (D,1,1) \\ 
 &+ 5 \times (L,1,1)+ 2 \times (L^c,1,1) + 3 \times(E^c,1,1)+ (L,\bar{5},1)+(L^c,\bar{5},1) ,
 \end{eqnarray*}
 where the chirality of the $L,L^c$ mirror pair is as indicated, ${\it i.e.}$ the pair is chiral with respect to
 $SU(5)$. 
 
 Standard model singlets:
 \begin{eqnarray*}
\\ &47  \times  (N,1,1)+ 10   \times  (N,5,1) + 2   \times  (N,\overline{10},1) + 7   \times  (N,1,16) + \\ &9 \times (N,1,10)
+ 3 \times (N,10,1) + (N,1,\overline{16}) + 3 \times (N,\bar5,1) 
 \end{eqnarray*}

Fractionally charged matter that is vector-like with respect to the standard model and $SU(5)_{B-L}$:
 \begin{eqnarray*}
             & (3,1,\frac13,1,1) + 
              (1,1,-\frac13,5,1) + 
              (1,1,\frac13,5,1) + 
              \\
&  
              (1,1,-\frac23,5,1) + 
              (1,1,\frac23,5,1) +
               \\
 &
 10 \times            (1,1,\frac13,1,1) + 
            (1,1,\frac13,1,10) + 
              \\
   &
   6 \times            (1,1,\frac23,1,1) + 
 14 \times            (1,2,\frac16,1,1) + 
  \\
  & 
  2 \times            (3,1,0,1,1) + 
   2 \times            (3,2,-\frac12,1,1)   +  {\rm c.c}
 \end{eqnarray*}
 
Fractionally charged matter that is vector-like with respect to the standard model but not with respect to $SU(5)_{B-L}$:
 \begin{eqnarray*}
              (1,2,\frac16,\bar{5},1) + 
              (1,2,-\frac16,\bar{5},1)   \\
 \end{eqnarray*}
Note that the chirality of the latter states, as that of the $(L,L^c)$ mirror pair listed above, originates entirely from 
$SU(5)$. The total chiral $SU(5)$ representation content is $11 (\bar 5)+10(5)+3(10)+2(\overline{10})$, which is indeed
anomaly free.  Phenomenologically this model is more problematic than the half-integer charge example presented above, especially
because of the $L, L^c$ mirror pair that couples chirally to $SU(5)_{B-L}$. 

For the  type SM $Q=\frac16$ we did not find any chiral spectra with $N$ families, $N>0$. There are non-chiral ({\it i.e.} $N=0$) spectra of this
type, and spectra with chiral exotics, for which $N$ is not defined.

\section{Conclusions}

Using B-L lifting we have found another large set of areas in the heterotic interacting CFT landscape 
where three family models
are easy to obtain, and are not strongly suppressed with respect to two or four family models. In comparison to heterotic
weight lifting in the internal sector,
models of this type should be more easily accessible by other methods since the internal $c=9$ CFT is not modified. On the
other hand, a ${\bf Z}_5$ simple current symmetry of the internal CFT is required to get chiral spectra, which limits the
applicability. 

A second way of understanding the two lifts of B-L is in terms of different embeddings of 
$SU(5)$ in the four-dimensional heterotic string gauge groups $SO(10)\times E_8$ or
$SO(26)$. Ironically, the canonical embedding of $SU(5)$ in $SO(10)$, the standard Gepner model case, yields
an even number of families, with only one exception \cite{Gepner:1987hi}. The counter-intuitive embedding in the extra $E_8$ (normally playing the
r\^ole of a ``hidden sector") is the one that produces three families in abundance. 

In field-theoretic constructions the embedding of $SU(5)$ in $E_8$ would of course never give rise to a chiral spectrum.
The reason this intuition is wrong is precisely that heterotic string constructions do not work like group-theoretic embeddings, as
already observed by Witten in \cite{Witten:1985xc}. This fact implies a second irony: the fact that heterotic GUTs, despite their
GUT gauge group, do not automatically have integrally charged spectra, unlike field theory GUTs. 

These two ironies cast a shadow on the apparent success of heterotic strings in getting $SO(10)$ GUTs almost
automatically. Even though this was the starting point of our work, it is difficult to argue that the observed standard model, with three
families and no fractional charges follows naturally from that starting point. But from the phenomenological point of view, the optimistic conclusion is that at least three family models are now very easy to obtain.

\vskip 2.truecm
\noindent
{\bf Acknowledgements:}
\vskip .2in
\noindent
This paper is dedicated to the memory of Max Kreuzer, who passed away while we were finishing it.
Max will live on through his work, and this paper is an example. Indeed, the formalism developed in 
\cite{Kreuzer:1993tf} is one of the corner stones of the present work.
\noindent
\leftline{ }
\noindent
This work has been partially 
supported by funding of the Spanish Ministerio de Ciencia e Innovaci\'on, Research Project
FPA2008-02968, and by the Project CONSOLIDER-INGENIO 2010, Programme CPAN
(CSD2007-00042). The work of A.N.S. has been performed as part of the program
FP 57 of Dutch Foundation for Fundamental Research of Matter (FOM). 

\bibliography{REFS}
\bibliographystyle{lennaert}

\end{document}